\documentclass[twocolumn,amsmath,amssymb]{revtex4}
\usepackage[dvips]{graphicx}
\usepackage{dcolumn}
\usepackage{bm}

\begin{document}
\title[]
{The stability of thin-shell wormholes with a phantom-like
     equation of state}
\author{Peter K.\,F. Kuhfittig}
\email{kuhfitti@msoe.edu}
\address{Department of Mathematics\\
Milwaukee School of Engineering\\
Milwaukee, Wisconsin 53202-3109 USA}

\begin{abstract}\noindent
This paper discusses the stability to linearized radial perturbations
of spherically symmetric thin-shell wormholes with a ``phantom-like"
equation of state for the exotic matter at the throat: $\mathcal{P}=
\omega\sigma$, $\omega<0$, where $\sigma$ is the energy-density of the
shell and $\mathcal{P}$ the surface pressure.  This equation is
analogous to the generalized Chaplygin-gas equation of state used by
E.F. Eiroa.  The analysis, which differs from Eiroa's in its basic
approach, is carried out for wormholes constructed from the following
spacetimes: Schwarzschild, de Sitter and anti de Sitter,
Reissner-Nordstr\"{o}m, and regular charged black-hole spacetimes,
as well as from black holes in dilaton and generalized dilaton-axion
gravity. \\
\phantom{a}\\
PAC numbers: 04.20.Jb, 04.20.Gz
\end{abstract}

\maketitle
\noindent

\section{Introduction}\noindent
A powerful theoretical method for describing or mathematically
constructing a class of spherically symmetric wormholes
from black-hole spacetimes was proposed by Visser in
1989 \cite{mV89}.  This type of wormhole, constructed by the
so-called cut-and-paste technique, is commonly known as a
\emph{thin-shell wormhole}, since the construction calls
for grafting two black-hole spacetimes together.  The
junction surface is a three-dimensional thin shell.  The
cut-and-paste technique is now considered standard.

While there had already been a number of forerunners, the
concept of a traversable wormhole was proposed by Morris
and Thorne in 1988 \cite{MT88}.  Ten years later a
renewed interest was sparked by the discovery that our
Universe is undergoing an accelerated expansion \cite
{aR98, sP99}: $\overset{..}{a}>0$ in the Friedmann
equation $\overset{..}{a}/a=-\frac{4\pi}{3}(\rho+3p)$.
(Our units are taken to be those in which $G=c=1$.)
The acceleration is caused by a negative pressure
\emph{dark energy} with equation of state (EoS)
$p=\omega\rho$, $\omega<-\frac{1}{3}$, and $\rho>0$.  A
value of $\omega<-\frac{1}{3}$ is required for an
accelerated expansion, while $\omega=-1$ corresponds to
a cosmological constant \cite{mC01}.  The case
$\omega<-1$ is referred to as \emph{phantom energy}
and leads to a violation of the null energy
condition, a primary prerequisite for the existence
of wormholes.  Wormholes may also be supported by
a generalized Chaplygin gas \cite {fL06} whose EoS
is $p=-A/\rho^{\alpha}$, where $A>0$ and
$0<\alpha\le 1$.

In a thin-shell wormhole the exotic matter is confined
to the thin shell.  This suggests assigning an equation
of state to the exotic matter on the shell.  Eiroa
\cite{eE09} used the above generalized Chaplygin EoS
$\mathcal{P}=A/|\sigma|^{\alpha}$, where $\sigma$ is the
(negative) energy-density of the shell and $\mathcal{P}$
the surface presssure, to perform a stability analysis
for linearized radial perturbations.  In this paper we
will consider, analogously, the EoS
$\mathcal{P}=\omega\sigma$, $\omega<0$, which will be
called a \emph{phantom-like} equation of state.  The
stability analysis will be carried out for several
spacetimes: Schwarzschild, de Sitter and anti de
Sitter, Reissner-Nordstr\"{o}m, and regular charged
black hole spacetimes, as well as black holes in
dilaton and generalized dilaton-axion gravity.  The
phantom-like equation of state yields explicit
closed-form expressions for $\sigma$.  Our approach
to the stability analysis is therefore different
from Eiroa's.

\section{Thin-shell wormhole construction}\noindent
Our starting point is the spherically symmetric metric
\cite {eE09}
\begin{equation}\label{E:line}
ds^2 = -f(r) dt^2 + [f(r)]^{-1}dr^2
  + h(r)(d\theta^2+\sin^2\theta d\phi^2),
\end{equation}
where $f(r)$ and $h(r)$ are positive functions of $r$
and $h(r)$ is increasing.  (In Sections III-VI,
$h(r)=r^2$.)  As in Ref. \cite{PV95}, the construction
begins with two copies of a black-hole spacetime and
removing from each the four-dimensional region
\begin{equation}\label{E:remove}
  \Omega^\pm = \{r\leq a\,|\,a>r_h\},
\end{equation}
where $r=r_h$ is the (outer) event horizon of the black
hole.  Now identify (in the sense of topology) the
time-like hypersurfaces
\[
  \partial\Omega^\pm =\{r=a\,|\,a>r_h\}.
\]
The resulting manifold is geodesically complete and
possesses two asymptotically flat regions connected
by a throat.  Next, we use the Lanczos equations
\cite{mV89, eE09, PV95, LC04, ER04, TSE06, RKC06, RKC07,
RS07, LL08}
\begin{equation}\label{E:Lanczos}
  S^i_{\phantom{i}j}=-\frac{1}{8\pi}\left([K^i_{\phantom{i}j}]
   -\delta^i_{\phantom{i}j}[K]\right),
\end{equation}
where $[K_{ij}]=K^{+}_{ij}-K^{-}_{ij}$ and $[K]$ is the
trace of $K^i_{\phantom{i}j}$.  In terms of the surface
energy-density $\sigma$ and the surface pressure
$\mathcal{P}$, $S^i_{\phantom{i}j}=\text{diag}(-\sigma,
\mathcal{P}, \mathcal{P})$.  The Lanczos equations now yield
\begin{equation}\label{E:sigma}
  \sigma=-\frac{1}{4\pi}[K^\theta_{\phantom{\theta}\theta}]
\end{equation}
and
\begin{equation}\label{E:LanczosP}
  \mathcal{P}=\frac{1}{8\pi}\left([K^\tau_{\phantom{\tau}\tau}]
    +[K^\theta_{\phantom{\theta}\theta}]\right).
\end{equation}

A dynamic analysis can be obtained by letting the radius $r=a$
be a function of time \cite{PV95}.  As a result,
\begin{equation}\label{E:sigma}
\sigma = - \frac{1}{2\pi a}\sqrt{f(a) + \dot{a}^2}
\end{equation}
and
\begin{equation}\label{E:P}
  \mathcal{P} =  -\frac{1}{2}\sigma + \frac{1}{8\pi
}\frac{2\ddot{a} + f^\prime(a) }{\sqrt{f(a) + \dot{a}^2}}.
\end{equation}
Since $\sigma$ is negative on the shell, we are dealing with
exotic matter.  In fact, the weak energy condition (WEC) is
trivially satisfied since the radial pressure $p$ is zero
for a thin shell.  (The WEC requires the stress-energy
tensor $T_{\alpha\beta}$ to obey $T_{\alpha\beta}
\mu^{\alpha}\mu^{\beta}\ge 0$ for all time-like vectors and,
by continuity, all null vectors.)  So for the radial outgoing
null vector $(1,1,0,0)$, we therefore have $T_{\alpha\beta}
\mu^{\alpha}\mu^{\beta}=\rho+p=\sigma +0<0$.

\section{Schwarzschild wormholes}\label{S:Schwarzschild}\noindent
For our first case, the Schwarzschild spacetime, $h(r)=r^2$ in
line element (\ref{E:line}), as noted earlier.  Also, recall
that the radius $r=a$ is a function of time.  It is easy to
check that $\mathcal{P}$ and $\sigma$ obey the conservation equation
\[
   \frac{d}{d\tau}(\sigma a^2)+\mathcal{P}\frac{d}{d\tau}(a^2)=0.
\]
(In Eqs. (\ref{E:sigma}) and (\ref{E:P}), the overdot
denotes the derivative with respect to $\tau$.)  The equation
can be written in the form
\begin{equation}\label{E:conservation}
  \frac{d\sigma}{da} + \frac{2}{a}(\sigma+\mathcal{P}) = 0.
               \end{equation}

For a static configuration of radius $a_0$, we have
$\dot{a}=0$ and $\ddot{a}=0$.  Moreover, we will consider
linearized fluctuations around a static solution
characterized by the constants $a_0$, $\sigma_0$, and
$\mathcal{P}_0$.  Given the EoS $\mathcal{P}=\omega\sigma$,
Eq. (\ref{E:conservation}) can be solved by separation of
variables to yield
\begin{equation*}
  |\sigma(a)|=|\sigma_0|\left(\frac{a_0}{a}
    \right)^{2(\omega+1)},
\end{equation*}
where $\sigma_0=\sigma(a_0)$.  So the solution is
\begin{equation}\label{E:sigmaexplicit}
  \sigma(a)=\sigma_0\left(\frac{a_0}{a}
  \right)^{2(\omega+1)},\quad \sigma_0=\sigma(a_0).
\end{equation}

Next, we rearrange Eq. (\ref{E:sigma}) to obtain the
equation of motion
\begin{equation*}
\dot{a}^2 + V(a)= 0.
\end{equation*}
Here the potential $V(a)$ is defined  as
\begin{equation}\label{E:Vdefined}
V(a) =  f(a) - \left[2\pi a \sigma(a)\right]^2.
\end{equation}
Expanding $V(a)$ around $a_0$, we obtain
\begin{eqnarray}
V(a) &=&  V(a_0) + V^\prime(a_0) ( a - a_0) +
\frac{1}{2} V^{\prime\prime}(a_0) ( a - a_0)^2  \nonumber \\
&\;& + O\left[( a - a_0)^3\right].
\end{eqnarray}
Since we are linearizing around $a=a_0$, we require that
$V(a_0)=0$ and $V'(a_0)=0$.  The configuration is in
stable equilibrium if $V''(a_0)>0$.

Now recall that for the Schwarzschild spacetime,
$f(r)=1-2M/r$.  It follows that
\begin{multline*}
  V(a)=1-\frac{2M}{a}-4\pi^2a^2\sigma^2\\
   =1-\frac{2M}{a}-4\pi^2a^2\sigma_0^2
     \left(\frac{a_0}{a}\right)^{4+4\omega}
\end{multline*}
from Eq. (\ref{E:sigmaexplicit}).  From Eq.
(\ref{E:sigma}) with $\dot{a}=0$,
\[
 \sigma_0=-\frac{1}{2\pi a_0}\sqrt{1-\frac{2M}{a_0}},
\]
so that
\begin{equation}\label{E:VSch}
  V(a)=1-\frac{2M}{a}-\left(1-\frac{2M}{a_0}\right)
  \frac{a_0^{2+4\omega}}{a^{2+4\omega}}.
\end{equation}
The first requirement, $V(a_0)=0$, is clearly met,
but not the second.  (If the exotic matter on the
shell were not required to meet the extra condition
in the form of an EoS, then $V'(a_0)$ would indeed
be zero \cite{PV95}.)  From
\begin{multline*}
   V'(a_0)=\\ \frac{2M}{a_0^2}-\left(1-\frac{2M}
   {a_0}\right)a_0^{2+4\omega}(-2-4\omega)
       a_0^{-3-4\omega}=0,
\end{multline*}
we obtain the condition
\begin{equation}\label{E:omegaSch}
   \omega=-\frac{1}{2}\frac{a_0/M-1}{a_0/M-2}.
\end{equation}
Observe that as $a_0\rightarrow+\infty$, $\omega
\rightarrow -1/2-$, and as $a_0\rightarrow 2M+$,
$\omega\rightarrow -\infty$.  At $a_0=3M$, $\omega=-1$.
Substituting in
\begin{multline*}
  V''(a)=\\-\frac{4M}{a^3}-\left(1-\frac{2M}{a_0}\right)
   a_0^{2+4\omega}(2+4\omega)(3+4\omega)a^{-4-4\omega}
\end{multline*}
and simplifying, we obtain the intermediate result
\begin{equation}\label{E:UPSch}
  V''(a_0)=\frac{2}{a_0^2}\left(-\frac{2}{a_0/M}
      +\frac{1}{a_0/M}
   \frac{a_0/M-4}{a_0/M-2}\right)>0.
\end{equation}
Since the Schwarzschild black hole has an event horizon
at $r=2M$, $a_0/M-2>0$, and we conclude that the
inequality $V''(a_0)>0$ can only be satisfied if
\[
      a_0<0.
\]
As a result, there are no stable solutions.

To allow a comparison to some of the other cases, let
us choose (arbitrarily) $a_0/M=5$, as a result of which
$\omega=-2/3$, and plot $V(a)$ against $a/M$, as shown
in Fig. 1.
\begin{figure}[htbp]
\begin{center}
\includegraphics [clip=true, draft=false, bb= 0 0 305 190,
angle=0, width=4.5 in,
height=2.5 in, viewport=40 40 302 185]{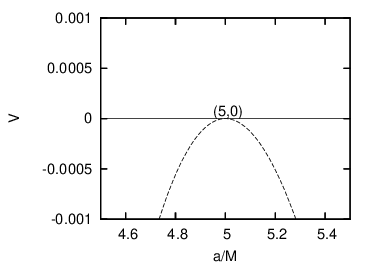}
\end{center}
\caption{The wormhole is unstable.}
\end{figure}

The more general analysis in Ref. \cite{PV95} depends
on the parameter $\beta^2(\sigma)=
\partial\mathcal{P}/\partial\sigma$, where $\beta$ is
usually interpreted as the speed of sound, so that
$0<\beta^2\le 1$.  There are no stable solutions in this
range.  However, as discussed in Ref. \cite{PV95}, since
we are dealing with exotic matter, this assumption may
be questioned, that is, $\beta^2$ may be just a
convenient parameter.  In that case, some stable
configurations may not be out of question.  Our
additional assumption, the EoS $\mathcal{P}=\omega\sigma$
on the shell, eliminates this possibility.

\section{Wormholes with a cosmological constant}
    \label{S:Lambda}\noindent
\subsection{Schwarzschild-de Sitter spacetimes}\noindent
In  the presence of a cosmological constant, $f(r)=
1-\frac{2M}{r}-\frac{1}{3}\Lambda r^2$.  For the de Sitter
case, $\Lambda>0$.  To keep $f(r)$ from becoming negative,
we must have $\Lambda M^2\le 1/9$.  This condition results
in two event horizons, where the inner horizon is between
$2M$ and $3M$.  (See Ref. \cite{eE09} for details.)  We
therefore assume that $a$ is greater than the outer
horizon.  Proceeding as in Sec. \ref{S:Schwarzschild},
\begin{multline}\label{E:VdeSitter}
  V(a)=1-\frac{2M}{a}\\
   -\frac{1}{3}\Lambda a^2-\left(1-\frac{2M}{a_0}
   -\frac{1}{3}\Lambda a_0^2\right)\left(
    \frac{a_0}{a}\right)^{2+4\omega}.
\end{multline}
Observe that $V(a_0)=0$.  As before, we have to determine
the condition on $\omega$ so that $V'(a_0)=0$:
\begin{equation}\label{E:omegadeSitter}
   \omega=-\frac{1}{2}\frac{1-1/(a_0/M)-(2/3)\Lambda
   M^2(a_0/M)^2}{1-2/(a_0/M)-(1/3)\Lambda M^2(a_0/M)^2}.
\end{equation}
(As in the Schwarzschild case, as $a_0\rightarrow
+\infty$, $\omega\rightarrow -1/2-$, and $\omega
\rightarrow-\infty$ as $a_0$ approaches the outer
event horizon.)   Substituting in $V''(a_0)$ and
simplifying, we get
\begin{multline}\label{E:UPdeSitter}
   V''(a_0)=\\ \frac{2}{a_0^2}\frac{-1/(a_0/M)
    +3\Lambda M^2(a_0/M)-
   (2/3)\Lambda M^2(a_0/M)^2}{1-2/(a_0/M)-(1/3)
      \Lambda M^2(a_0/M)^2}\\>0.
\end{multline}
The form of $V''(a_0)$ forces us to consider two cases,
a positive and negative denominator.

If the denominator is positive, then
\begin{equation}\label{E:Lambdagreater}
   \Lambda M^2>\frac{1}{(a_0/M)[3(a_0/M)-(2/3)(a_0/M)^2]}.
\end{equation}
This inequality implies that $a_0/M<4.5$ to keep the right
side positive.  It is easy to show analytically that
$\Lambda M^2>1/9$; in fact, $(3, 1/9)$ is a minimum.
We may also plot $\Lambda M^2$ against $a/M$, as
shown in Fig. 2.
\begin{figure}[htbp]
\begin{center}
\includegraphics [clip=true, draft=false, bb= 0 0 305 190,
angle=0, width=4.5 in,
height=2.5 in, viewport=40 40 302 185]{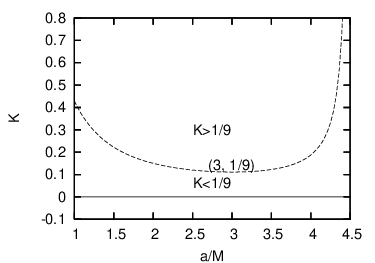}
\end{center}
\caption{$K=\Lambda M^2$ is plotted against $a/M$.}
\end{figure}
So for this case, the condition $V''(a_0)>0$ cannot be met
(since we must have $\Lambda M^2\le 1/9$), and we get
only unstable solutions.  Plotting $V(a)$ around
$a_0/M=5$ yields a graph that is very similar to the
graph in Fig. 1.

For the second case,
\begin{equation}\label{E:Lambdaless}
  1-\frac{2}{a_0/M}-\frac{1}{3}\Lambda M^2\left(
    \frac{a_0}{M}\right)^2<0
\end{equation}
in inequality (\ref{E:UPdeSitter}) we obtain
\begin{equation}\label{E:Lambdafirst}
  \Lambda M^2\left[3\left(\frac{a_0}{M}\right)-
   \frac{2}{3}\left(\frac{a_0}{M}\right)^2
       \right]<\frac{1}{a_0/M}.
\end{equation}
If $a_0/M>4.5$, then the left side is negative, and the
condition is automatically satisfied.  If $a_0/M<4.5$,
then, according to Fig. 2, $\Lambda M^2<1/9$, the
region below the graph.  So we conclude that in the
second case, the wormholes are stable.

For comparison, let us choose $a_0/M=5$ again and
$\Lambda M^2=0.11<1/9$, resulting in $\omega=-1.63$.
The plot of $V(a)$ against $a/M$ is shown in Fig. 3.
\begin{figure}[htbp]
\begin{center}
\includegraphics [clip=true, draft=false, bb= 0 0 305 190,
angle=0, width=4.5 in,
height=2.5 in, viewport=40 40 302 185]{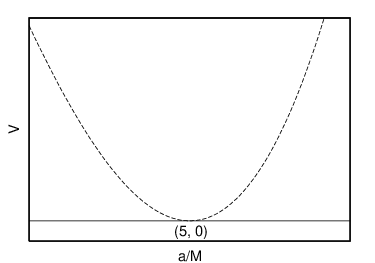}
\end{center}
\caption{The wormhole is stable.}
\end{figure}

In summary, in the Schwarzschild-de Sitter case, the
thin-shell wormholes are stable if, and only if,
\[
   1-\frac{2}{a_0/M}-\frac{1}{3}\Lambda M^2
     \left(\frac{a_0}{M}\right)^2<0.
\]

\subsection{Schwarzschild-anti de Sitter spacetimes}
\noindent
To study the case $\Lambda<0$, we return to inequality
(\ref{E:UPdeSitter}) and consider first a negative
denominator:
\[
    1-\frac{2}{a_0/M}-\frac{1}{3}\Lambda M^2
   \left(\frac{a_0}{M}\right)^2<0.
\]
Solving for $\Lambda M^2$, we obtain
\[
   \Lambda M^2>\frac{3-6/(a_0/M)}{(a_0/M)^2}.
\]
Since $a_0/M>2$, we conclude that $\Lambda M^2>0$, so
that this case cannot arise.

Reversing the sense of the inequality, we have from
inequlity (\ref{E:UPdeSitter})
\[
   \Lambda M^2\left[3\left(\frac{a_0}{M}\right)-
   \frac{2}{3}\left(\frac{a_0}{M}\right)^2
       \right]>\frac{1}{a_0/M}.
\]
Then the second factor on the left must be negative, which
implies that $a_0/M>4.5$.

So the wormhole is stable whenever
\begin{equation*}
  (1)\quad \Lambda M^2<\frac{1}{(a_0/M)[3(a_0/M)
     -(2/3)(a_0/M)^2]}
\end{equation*}
and
\begin{equation*}
  (2)\quad a_0/M>4.5.\phantom{wwwwwwwwwwwwwwwww}
\end{equation*}

\section{Reissner-Nordstr\"{o}m wormholes}
    \label{S:Reissner}\noindent
If the starting point is a Reissner-Nordstr\"{o}m
spacetime, then
\begin{equation}
   f(r)=1-\frac{2M}{r}+\frac{Q^2}{r^2},
\end{equation}
where $M$ and $Q$ are the mass and charge, respectively,
of the black hole.  For $0<|Q|<M$, this black hole has
two event horizons at $r=M\pm \sqrt{M^2-Q^2}$.  As
usual, we require that $r=a$ is larger than the outer
horizon.

Here we have
\begin{multline}\label{E:VReissner}
  V(a)=1-\frac{2M}{a}+\frac{Q^2}{a^2}\\
   -\left(1-\frac{2M}{a_0}+\frac{Q^2}{a_0^2}\right)
   \left(\frac{a_0}{a}\right)^{2+4\omega}.
\end{multline}
Once again, $V(a_0)=0$.  From $V'(a_0)=0$ we obtain
\begin{equation}\label{E:omegaReissner}
  \omega=-\frac{1}{2}\frac{(a_0/M)^2-a_0/M}
    {(a_0/M)^2-2(a_0/M)+Q^2/M^2}.
\end{equation}

Substituting into $V''(a_0)$ and simplifying, yields the
following inequality:
\begin{multline}\label{E:UPReissner}
  V''(a_0)=\\\frac{2}{a_0^2}
  \frac{-a_0/M-(Q^2/M^2)[1/(a_0/M)]+2Q^2/M^2}
   {(a_0/M)^2-2(a_0/M)+Q^2/M^2}>0.
\end{multline}
Since $a_0/M>2$, the denominator is positive.  Solving for
$Q^2/M^2$, leads to
\begin{equation}\label{E:Reissnercondition}
  \frac{|Q|}{M}>\frac{a_0/M}{\sqrt{2(a_0/M)-1}},
\end{equation}
which exceeds unity.  To meet this condition, $|Q|$
would have to exceed $M$.

So to obtain a stable solution, we will have to
tolerate a naked singularity at $r=0$, but since
$a_0>0$, the naked singularity is removed from
the wormhole spacetime.

\section{Wormholes from regular charged black holes}
\label{S:regular}\noindent
Thin-shell wormholes from regular charged black holes,
due to Ayon-Beato and Garc\'{i}a \cite {AG99}, are
discussed in Ref. \cite {RK09}.  For this black hole,
\begin{equation}\label{E:ftanh}
  f(r)=1-\frac{2M}{r}+\frac{2M}{r}\text{tanh}
   \left(\frac{Q^2}{2Mr}\right).
\end{equation}
Again, $M$ and $Q$ are the mass and charge,
respectively.  It is shown in Ref. \cite{AG99} that the
black hole has two event horizons whenever
$|Q|<1.05M$.  Consider next
\begin{multline}\label{E:Vregular}
   V(a)=1-\frac{2M}{a}+\frac{2M}{a}\text{tanh}
      \left(\frac{Q^2}{2Ma}\right)\\
   -\left[1-\frac{2M}{a_0}+\frac{2M}{a_0}\text{tanh}
   \left(\frac{Q^2}{2Ma_0}\right)\right]
   \left(\frac{a_0}{a}\right)^{2+4\omega}.
\end{multline}
As before, $V(a_0)=0$, and from $V'(a_0)=0$, we get
\begin{equation}\label{E:omegaregular}
  \omega=
  \frac{1}{2}\left[-1+\frac{g(a_0)}
  {a_0/M-2+2\,\text{tanh}[Q^2/(2Ma_0)]}\right],
\end{equation}
where
\begin{multline*}
  g(a_0)=\\-1+\text{tanh}\left(\frac{Q^2}{2Ma_0}\right)
   +\frac{Q^2/M^2}{2a_0/M}\text{sech}^2
    \left(\frac{Q^2}{2Ma_0}\right).
\end{multline*}

Based on the graphical output, we get only unstable
solutions.  For example, choosing $a_0/M=5$ again for
comparison and letting $|Q|/M=0.9$, we get $\omega=
-0.63$.  The resulting graph, shown in Fig. 4,
resembles Fig. 1.  Other choices of the parameters
\begin{figure}[htbp]
\begin{center}
\includegraphics [clip=true, draft=false, bb= 0 0 305 190,
angle=0, width=4.5 in,
height=2.5 in, viewport=40 40 302 185]{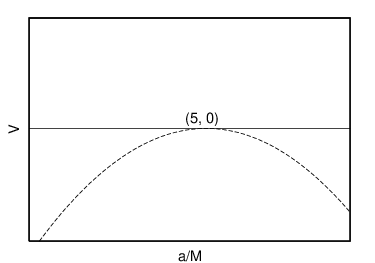}
\end{center}
\caption{The wormhole is unstable.}
\end{figure}
lead to similar results.

For completeness, let us note that a more general
analysis using the parameter $\beta^2(\sigma)=
\partial/\mathcal{P}/\partial\sigma$ instead of an
EoS is presented in Ref. \cite{RK09}: for any
$|Q|/M<1.05$, there exists a narrow range of values
for $a_0$ for which the wormhole is stable.

\section{Wormholes from black holes in dilaton and
dilaton-axion gravity}\noindent
Of the remaining thin-shell wormholes, based on dilaton
and dilaton-axion black holes, respectively, we will
consider in detail only the latter, which is the more
complicated of the two.

The dilaton-axion black-hole solution, inspired by
low-energy string theory, was discovered by Sur,
\emph{et al.}, \cite {SDS05} and is also discussed in
Ref. \cite{RK10}.  We need to list certain parameters
in order to define $V(a)$.  As in the
Reissner-Nordstr\"{o}m wormhole, there are two event
horizons, denoted by $r_{-}$ and $r_{+}$, respectively.
Returning now to line element (\ref{E:line}), we can
list both $f(r)$ and $h(r)$ \cite{SDS05}:
\[
   f(r)=\frac{(r-r_{-})(r-r_{+})}
    {(r-r_0)^{2-2n}(r+r_0)^{2n}};
\]
\[
   h(r)=\frac{(r+r_0)^{2n}}{(r-r_0)^{2n-2}}.
\]
Since $h(r)$ is no longer equal to $r^2$ in line element
(\ref{E:line}), Eq. (\ref{E:sigma}) becomes
\begin{equation}\label{E:sigmaaxion}
   \sigma=-\frac{1}{4\pi}\frac{h'(a)}{h(a)}
      \sqrt{f(a)+\dot{a}^2}
\end{equation}
and the conservation equation (\ref{E:conservation})
has to be replaced by \cite{eE09}:
\begin{multline}\label{E:conservationaxion1}
  \frac{d}{d\tau}(\sigma\mathcal{A})
     +\mathcal{P}\frac{d\mathcal{A}}{d\tau}=\\
  \{[h'(a)]^2-2h(a)h''(a)\}
  \frac{\dot{a}\sqrt{f(a)+\dot{a}^2}}{2h(a)},
\end{multline}
where $\mathcal{A}=4\pi h(a)$ is the area of the
throat by Eq. (\ref{E:line}).  The prime and dot
denote, respectively, the derivatives with
respect to $a$ and $\tau$.  Substituting
Eq. (\ref{E:sigmaaxion}) on the right-hand side,
we get
\begin{multline*}
   \frac{d}{d\tau}[4\pi h(a)\sigma]+\mathcal{P}
   \frac{d}{d\tau}[4\pi h(a)]\\
  =-\{[h'(a)]^2-2h(a)h''(a)\}
     \frac{\dot{a}(4\pi\sigma)}{2h'(a)},
\end{multline*}
whence
\begin{multline*}
  \frac{d}{da}[\sigma h(a)]+\mathcal{P}
  \frac{d}{da}[h(a)]\\
   =-\{[h'(a)]^2-2h(a)h''(a)\}\frac{\sigma}{2h'(a)}.
\end{multline*}
Our final form is
\begin{multline}\label{E:conservationaxion2}
  h(a)\sigma'+h'(a)(\sigma+\mathcal{P})\\
  +\{[h'(a)]^2-2h(a)h''(a)\}\frac{\sigma}{2h'(a)}.
\end{multline}
Making use of $\mathcal{P}=\omega\sigma$, this
equation can be solved by separation of variables:
\begin{equation}\label{E:sigmaaxion2}
  \sigma(a)=\sigma_0\left[\frac{h(a_0)}{h(a)}
     \right]^{3/2+\omega}
  \left[\frac{h'(a_0)}{h'(a)}\right]^{-1}.
\end{equation}
(Here we used the fact that $h'(a)>0$.)  It is
shown in Ref. \cite{RK10} that
\begin{equation}\label{E:sigmazero}
   \sigma_0=-\frac{4[a_0+(1-2n)r_0]
          (a_0-r_{-})(a_0-r_{+})}
   {D(a_0-r_0)(a_0+r_0)},
\end{equation}
where
\begin{equation}\label{E:D}
  D=8\pi(a_0-r_0)^{1-n}(a_0+r_0)^n
   \sqrt{(a_0-r_{-})(a_0-r_{+})}.
\end{equation}
Using the equation of motion $\dot{a}^2+V(a)=0$
once again, we get from Eq. (\ref{E:sigmaaxion}),
\begin{equation}\label{E:Vdilaton}
  V(a)=f(a)
     -\left[4\pi\frac{h(a)}{h'(a)}\sigma(a)\right]^2.
\end{equation}
Eq. (\ref{E:sigmaaxion2}) now yields
\begin{multline}\label{E:Vfinal}
  V(a)=\frac{(a-r_{-})(a-r_{+})}{(a-r_0)^{2-2n}
         (a+r_0)^{2n}}\\
  -\left[4\pi\frac{h(a)}{h'(a_0)}\sigma_0\right]^2
    \left[\frac{h(a_0)}{h'(a)}\right]^{3+2\omega}.
\end{multline}
While it is easy enough to check that $V(a_0)=0$, it is no
longer convenient to compute $\omega$ as a function of
the various parameters.  Plotting $V(a)$ against $a$ instead
of $a/M$, we can determine $\omega$ by trial and error:
$V(a)$ must be tangent to the $a$-axis at $a=a_0$,
where $V(a_0)=0$ automatically.  For example, if
$a_0=5$, $r_0=1$, $r_{-}=2$, $r_{+}=2.05$, and $n=0.8$,
then $\omega=-0.915$.  If $a_0=5$, $r_0=1$, $r_{-}=2$,
$r_{+}=3$, and $n=0.8$, then $\omega=-1.132$.  Reducing
$n$ to $0.6$ produces $\omega=-0.84$ in the first case
and $\omega=-1.041$ in the second.  In all cases the
graphs are concave down at $a_0=5$ and look similar
to the graph in Fig. 1.  Based on the graphical output,
there do not appear to be any stable solutions.

For the dilaton case we have \cite{eE08}
\begin{multline}\label{E:Vdilaton}
  V(a)=\left(1-\frac{A}{a}\right)
   \left(1-\frac{B}{a}\right)^{(1-b^2)/(1+b^2)}\\
  -\left[4\pi\frac{h(a)}{h'(a)}\sigma_0\right]^2
    \left[\frac{h(a_0)}{h(a)}\right]^{3+2\omega},
\end{multline}
where $h(a)=a^2(1-B/a)^{2b^2/(1+b^2)}$ for various
constants.  Once again, one can readily check that
$V(a_0)=0$.

As in the dilaton-axion case, $\omega$ can be found
by trial and error.  For example, if $a_0=5$, $b=0.5$,
$A=2$, and $B=1$, then $\omega=-0.693$; if $a_0=6$,
$b=0.8$, $A=4$, and $B=2$, then $\omega=-0.94$, etc.
The resulting graphs are similar to those in the
dilaton-axion case.
\\
\\
\section{Conclusion}\noindent
This paper discusses the stability to linearized
radial perturbations of spherically symmetric thin-shell
wormholes with the equation of state $\mathcal{P}
=\omega\sigma$, $\omega<0$, for the exotic matter
at the throat.  This EoS is referred to as
phantom-like.  Various spacetimes were considered.

It was found that the wormholes are unstable if
constructed from Schwarzschild spacetimes or regular
charged black holes, as well as from black holes in
dilaton and dilaton-axion gravity.  For the
Reissner-Nordstr\"{o}m case, stable solutions exist
only if
\[
   \frac{|Q|}{M}>\frac{a_0/M}{\sqrt{2(a_0/M)-1}},
\]
leading to a naked singularity, which is, however,
removed from the wormhole spacetime.  For the
Schwarzschild-de Sitter case, the wormholes are
stable if, and only if,
\[
   1-\frac{2}{a_0/M}-\frac{1}{3}\Lambda M^2
        \left(\frac{a_0}{M}\right)^2<0.
\]
In the Schwarzschild-anti de Sitter case, the
configurations are stable whenever
\begin{equation*}
  (1)\quad \Lambda M^2<\frac{1}{(a_0/M)[3(a_0/M)-(2/3)(a_0/M)^2]}
\end{equation*}
and
\begin{equation*}
  (2)\quad a_0/M>4.5.\phantom{wwwwwwwwwwwwwwwww}
\end{equation*}

\end{document}